\documentclass[aps,pre,reprint,onecolumn,notitlepage]{revtex4-1}
\usepackage[english]{babel}
\usepackage{amsmath, amsthm, amssymb, latexsym, graphicx,lipsum}
\usepackage[utf8]{inputenc}
\usepackage{color}

\newcommand{\ua}{\textbf{v}_{\text{A}}}
\newcommand{\da}{\boldsymbol\nabla\cdot\textbf{v}_{\text{A}}}
\newcommand{\dap}{\boldsymbol\nabla'\cdot\textbf{v}_{\text{A}}'}
\newcommand{\uh}{\textbf{v}}
\newcommand{\dv}{\boldsymbol\nabla\cdot\textbf{v}}
\newcommand{\dvp}{\boldsymbol\nabla'\cdot\textbf{v}'}

\begin{document}

\title{Alternative derivation of exact law for compressible and isothermal magnetohydrodynamics turbulence}

\author{N. Andr\'es$^{1}$}
\author{F. Sahraoui$^{1}$}
\affiliation{$^1$ LPP, CNRS, Ecole Polytechnique, UPMC Univ. Paris 06, Univ. Paris-Sud, Observatoire de Paris, Université Paris-Saclay, Sorbonne Universités, PSL Research University, F-91128 Palaiseau, France}
\date{\today}

\begin{abstract}
The exact law for fully developed homogeneous compressible magnetohydrodynamics (CMHD) turbulence is derived. For an isothermal plasma, without the assumption of isotropy, the exact law is expressed as a function of the plasma velocity field, the compressible Alfv\'en velocity and the scalar density, instead of the  Elsasser variables used in previous works. The theoretical results show four different types of terms that are involved in the nonlinear  cascade of the total energy in the inertial range. Each category is examined in detail, in particular those that can be written either as source or flux terms. Finally, the role of the background magnetic field ${\bf B}_0$ is highlighted and comparison with the incompressible MHD (IMHD) model is discussed. This point is particularly important when testing the exact law on numerical simulations and {\it in situ} observations in space  plasmas.
\end{abstract}

\maketitle

\section{Introduction}\label{intro}

Exact laws for incompressible turbulence result from the so-called von-K\'arm\'an-Howarth (vKH) equation \citep{vkh1938} and represent one of the cornerstones of turbulence theories \citep{F1995}. The vKH dynamical equation relates the second-order correlation energy function with a third-order correlation function of the turbulent fields, which represent the energy flux through different scales in the system. Under the assumptions of {infinite kinetic Reynolds number, time stationarity, space homogeneity and full isotropy}, \citet{vkh1938} derived the first exact law for hydrodynamic (HD) turbulence valid in the inertial range, the so-called 4/5 law. This exact law predicts a linear scaling for the longitudinal third-order structure function of the velocity field with the distance between points. It also links information that is accessible at large scales, generally given by the structures functions of the turbulent fields, to small-scale quantities such as the energy dissipation rate. Among several uses, the vKH equation provides a precise identification of the inertial range \citep[see, e.g. ][and references therein]{MY1975}, an estimation of the energy cascade rate and the Reynolds numbers in turbulent experiments, in particular when dissipative mechanisms are unknown such as in near-Earth space plasmas \citep[see, e.g.][]{SV2007,WEY2007}.

{Deriving exact laws in fully developed turbulence that are valid in the inertial range requires introducing important assumptions, which are not always fulfilled in real turbulence experiments. These assumptions are statistical stationarity and homogeneity of turbulence, and finite energy cascade rate $\varepsilon$ as the viscosity (and/or magnetic resistivity) tend to zero. This last condition is particularly important since it allows one to compute energy cascade rates even when dissipation mechanisms (presumably acting in the smallest scales of the system) are negligible. Additionally, full spatial isotropy can be assumed. However, in the presence of a strong magnetic field in the plasma, the latter condition need to be relaxed \citep[see,][]{P2003,A2016c}.}

\citet{Ga2011} have reported the first derivation of the vKH equation for compressible HD turbulence within the isothermal approximation, revealing a major difference with respect to the incompressible case. The authors have indeed shown the presence of a new terms that act in the inertial range as a source (or a sink) for the mean energy cascade rate, while in incompressible HD turbulence there is only one type of term, the so-called flux terms that act to transfer energy in the inertial range \citep{vkh1938,K1941a,K1941b}. \citet{K2013} used three-dimensional (3D) numerical simulations of supersonic isothermal turbulence to validate the exact relation derived in \citet{Ga2011}. The authors found that the source terms are negligible with respect to the flux terms in the inertial range. They also showed that two compressible analogues of the 4/5 law exist describing fifth- and fourth-order correlations, but only the fourth-order relation remains ``universal'' in a wide range of Mach numbers covering nearly incompressible and highly compressible turbulent flows. Finally, \citet{B2014} investigated compressible HD turbulence under the assumption of a polytropic closure. The authors derived an exact relation and found that the inertial range is characterized by a flux term that is related to the enthalpy, and a purely compressible term that may act as a source (or a sink) for the mean energy transfer rate.

Several attempts have been made to extend the incompressible HD formalism to plasma turbulence \citep{Ch1951,P1998a,P1998b,Ch1996,Ga2008,B2016b,A2016b,A2016c}. In particular, \citet{Ch1951} derived the vKH equation for IMHD turbulence under the homogeneity and full isotropy assumptions. Assuming also equipartition between kinetic and magnetic energy, \citet{P1998a,P1998b} derived the 4/5 law for the second-order correlation energy functions for IMHD turbulence. In the absence of mirror symmetry, \citet{P2003} presented an exact equation for homogeneous and isotropic IMHD turbulence with nonzero helicity. In the large Reynolds number and long-time limit, the authors recovered a linear scaling for the third-order correlation tensors. The validity of the IMHD exact law has been subjected to several numerical tests \citep[see, e.g.][]{Mi2009,Bo2009,W2010}, and it has been used to estimate the energy cascade rate {\citep{SV2007,Sa2008,Va2007,M2008,C2009}} and the magnetic and kinetic Reynolds numbers \citep{WEY2007} in solar wind turbulence, and in large scale modeling of the solar wind \citep{M1999,Mc2008}.

The first generalization to CMHD turbulence has been reported in \citet{B2013}. Under the assumption of statistical homogeneity and in the infinite Reynolds numbers limit, the author derived an exact relation for two-point correlation functions of the turbulent fields. Using the compressible Elsasser variables \citep{M1987}, the authors reported a valid expression for the exact law for CMHD turbulence. However, the exact law was expressed in terms of flux or source terms only. In this paper, we revisit the original work of \citet{B2013} and provide a new derivation using the classical plasma variables instead of the Elsasser variables, which are not the most appropriate ones for compressible turbulence (their use in \citet{B2013} was motivated by the idea of direct comparison with the IMHD model of \citet{P1998a,P1998b}~\citep{B2017}). We show in particular that plasma compressibility brings three more types of terms with respect to the incompressible case, whose impact on the cascade of the total compressible energy is discussed.

From the observational viewpoint, the first attempt to include the role of density fluctuations in estimating the energy cascade rate in the solar wind has been reported in \citet{C2009}. The authors used Ulysses observations and a heuristic model to calculate a scaling relation of the energy flux. They found a significant increase in the turbulent cascade rate and a better scaling with respect to the incompressible exact law \citep{P1998a,P1998b}. However, those results were based on a heuristic model (and not on an exact law) and used pseudo-energies, which are not conserved in CMHD theory \citep{M1987}. Moreover, the level of enhancement given by the heuristic model has been partly questioned by \citet{H2017} who showed that it can be caused by the plasma mean flow that is involved in the model. Using exact law of \citet{B2013} and {\it in situ} data from the THEMIS spacecraft~\citep{Au2009}, \citet{B2016a} and \citet{H2017} have studied the role of compressibility in the energy cascade of the solar wind turbulence. The authors found a more prominent role of density fluctuations in amplifying the energy cascade rate in the slow than in the fast solar wind. Another interesting feature that has been evidenced in the terrestrial magnetosheath, which has higher density fluctuations ($\delta \rho/\rho\sim 50\%-100\%$) than the solar wind ($\delta \rho/\rho\sim 5\%-20\%$), is that density fluctuations reinforce the anisotropy of the energy cascade rate with respect to the local magnetic field~\citep{H2017b}.  Note however that in these recent works, only the flux terms were estimated while the source terms could not have been evaluated reliably using single spacecraft data because of their dependence on the local spatial derivatives. They were assumed to be sub-dominant in the inertial range based on HD numerical results \citep{K2013}.

The main goal of the present paper is to derive in a simpler way the exact law for CMHD turbulence and present a comprehensive analysis of  this law, with particular emphasis of the nature of each term involved in the nonlinear cascade of energy and of the role of the background magnetic field ${\bf B}_0$. {We express our results as function of the primitive variables, i.e., the density, the velocity and the compressible Alfv\'en velocity fields, which prove to be more suitable in CMHD theory than the classical Elsasser variables. Furthermore, we discuss the nature of each type of term present in the exact law, which helps clarifying some subtle issues regarding the use of the new compressible exact law on numerical simulations or spacecraft data and the comparison with the results from the IMHD model. The paper is organized as follows:  in Section \ref{equations} we briefly describe the CMHD set of equations. In Section \ref{derivation} we present a full derivation of the exact law in CMHD turbulence using the primitive variables instead of the Elsasser variables}.  In Section \ref{discussion} we discuss our main results and their implications on observational studies in the near-Earth space and on numerical simulations of CMHD. Finally, in Section \ref{conclusions} we summarize our main findings.

\section{Compressible magnetohydrodynamics equations}\label{equations}

The three-dimensional (3D) CMHD equations correspond to the momentum equation for the velocity field $\uh$, the induction equation for the magnetic field \textbf{B}, the continuity equation for the scalar density $\rho$ and the differential Gauss' law. These equations can be cast as \citep[e.g.][]{F2014,A2017a},
\begin{align}\label{modeld:1}
	\frac{\partial \textbf{v}}{\partial t} &= -\textbf{v}\cdot\boldsymbol\nabla\textbf{v}-\frac{\boldsymbol\nabla P}{\rho} + \frac{(\boldsymbol\nabla\times\textbf{B})\times\textbf{B}}{4\pi\rho} + \textbf{F}_k+\textbf{D}_k, \\ \label{modeld:2}
	\frac{\partial \textbf{B}}{\partial t} &= \boldsymbol\nabla\times\left(\textbf{v}\times\textbf{B}\right) + \textbf{D}_m, \\ \label{modeld:3}
	\frac{\partial \rho}{\partial t} &= -\boldsymbol\nabla\cdot(\textbf{v}\rho), \\ \label{modeld:4}
	\boldsymbol\nabla\cdot\textbf{B}&=0,
\end{align}
where \textbf{v} is the velocity field fluctuations, $\textbf{B}=\textbf{B}_0+\textbf{b}$ is the total magnetic field (${\bf B}_0$ is the background and ${\bf b}$ the fluctuating magnetic field), $\rho$ and $P$ are the scalar density and pressure, respectively. For the sake of simplicity we assume that the plasma obeys an isothermal equation of state, $P=c_s^2\rho$, where $c_s$ is the constant sound speed. {The use of the isothermal assumption is aimed at simplifying the complex analytical derivation of the exact law of compressible turbulence. While the isothermal closure may not be fully adequate for collisionless plasmas as those of the near-Earth space, it has nevertheless been used in solar wind modeling \citet{P1958}) and was shown to reflect some features of compressible flows in the solar corona based on observations of the effective adiabatic index \citep{V2011}. The present model can be improved in the future by considering more general closure equations, e.g. a polytropic relation, where the scalar pressure is no longer proportional to the density.} Finally, \textbf{F}$_k$ represents the mechanical large-scale forcing and $\textbf{D}_{k,m}$ are the small-scale kinetic and magnetic dissipative terms, respectively.

Alternatively to the magnetic field \textbf{B}, the compressible Alfv\'en velocity $\ua = \textbf{B}/\sqrt{4\pi\rho}$ can be used \citep{M1987}, where the time and space dependence enters through ${\bf B}$ {\it and} ${\rho}$. In this manner, both variables $\uh$ and $\ua$ are expressed in speed units. It is worth mentioning that, while {we assume the background flow speed to be zero}, i.e., ${\uh}_0=0$, the background Alfv\'en speed in non-zero, i.e., ${\ua}_0 = \textbf{B}_0/\sqrt{4\pi\rho_0}$. Using $\uh$, $\ua$ and $\rho$, Eqs. \eqref{modeld:1}-\eqref{modeld:4} can be cast as \citep{M1987},
\begin{align}\label{1}
	\frac{\partial \textbf{v}}{\partial t} &= -\uh\cdot\boldsymbol\nabla\uh  + \ua\cdot\boldsymbol\nabla\ua - \frac{1}{\rho}\boldsymbol\nabla(P+P_M) - \ua(\da) + \textbf{D}_k+\textbf{F}_k,\\ \label{2}
	\frac{\partial \ua}{\partial t} &= - \uh\cdot\boldsymbol\nabla\ua + \ua\cdot\boldsymbol\nabla\uh -\frac{\ua}{2}(\dv) + \textbf{D}_m, \\ \label{3}
	\frac{\partial e}{\partial t}  &= - \uh\cdot\boldsymbol\nabla e-c_s^2\boldsymbol\nabla\cdot\uh, \\ \label{4}
	\ua\cdot\boldsymbol\nabla\rho &= -2\rho(\da).
\end{align}
Here $P_M\equiv\rho u_\text{A}^2/2$ is the magnetic pressure, and \eqref{modeld:3} has been rewritten as function of the internal compressible energy for an isothermal plasma, i.e. $e\equiv c_s^2\log(\rho/\rho_0)$~\citep{G2016}. In the following section, Eqs. \eqref{1}-\eqref{4} will be used to derive an exact law for CMHD turbulence.

\section{An alternative derivation of the vKH law in CMHD turbulence}\label{derivation}

In CMHD theory, the {density of the} total energy $E(\textbf{x})$ and the density-weighted cross helicity $H(\textbf{x})$ can given by
\begin{align}\label{energy}
	E(\textbf{x}) =&~ \frac{\rho}{2}(\uh\cdot\uh+\ua\cdot\ua) + \rho e, \\ \label{cross}
	H(\textbf{x}) =&~ \rho(\uh\cdot\ua),
\end{align}
while the total energy {(i.e., the spatial integral of Eq.~\ref{energy})} is one of the ideal invariants, the density-weighted cross helicity is not. Both quantities are essential for the derivation of the exact law in CMHD turbulence. {It is worth mentioning that the usual cross magnetic helicity is an ideal invariant in CMHD with a barotropic closure.}
We define the two-point correlation function associated with the total energy by
\begin{align}
	R_E(\textbf{x},\textbf{x}') \equiv \frac{\rho}{2}(\uh\cdot\uh'+\ua\cdot\ua') + \rho e'.
\end{align}
where the prime denotes field evaluation at $\textbf{x}'=\textbf{x}+{\bf r}$ (being {\bf r} the displacement vector) and the angular bracket $\langle\cdot\rangle$ denotes an ensemble average. The properties of spatial homogeneity implies {(assuming ergodicity)} that the results of averaging over a large number of realizations can be obtained equally well by averaging over a large region of space for one realization \citep{Ba1953}. In particular, under homogeneity assumption, the correlation functions will depend only on the vector displacement {\bf r}.

For the exact law derivation, a dynamical equation for the correlator $\langle R_E+R_E'\rangle$ is a key ingredient, since it is for this correlator that we can derive an exact law valid in the inertial range. It is defined as
\begin{align}\label{dynamical}
	\partial_t \langle R_E + R_E'\rangle &= \frac{1}{2}\partial_t\langle\rho\uh\cdot\uh' + \rho'\uh'\cdot\uh\rangle + \frac{1}{2}\partial_t\langle\rho\ua\cdot\ua' + \rho'\ua'\cdot\ua\rangle + \partial_t\langle\rho e' + \rho'e\rangle.
\end{align}
In a previous derivation of the exact law for CMHD turbulence \citep{B2013}, the authors calculated a dynamical equation for a different correlator that is $\langle R_E \pm R_H+R_E' \pm R_H'\rangle$  ($R_H(\textbf{x},\textbf{x}') \equiv \rho(\uh\cdot\ua'+\ua\cdot\uh')/2$ being the density-weighted cross helicity correlator) and then summed both equations. Here we take a different path and calculate directly the dynamical equation $\partial_t\langle R_E+R_E'\rangle$.

Using Eqs. \eqref{1}-\eqref{4} (evaluated both at points $\textbf{x}$ and $\textbf{x}'$) and basic vector algebra properties, it is possible to calculate each term in Eq.~\eqref{dynamical},
\begin{align}\nonumber
	\partial_t(\rho\uh\cdot\uh') =&~ - \boldsymbol\nabla\cdot[(\uh\cdot\uh')\rho\uh] + \boldsymbol\nabla\cdot[(\ua\cdot\uh')\rho\ua] - \boldsymbol\nabla'\cdot[(\uh'\cdot\uh)\rho\uh'] + \boldsymbol\nabla'\cdot[(\ua'\cdot\uh)\rho\ua']  \\ \nonumber
	& -\boldsymbol\nabla\cdot(P\uh') - \boldsymbol\nabla\cdot(P_M\uh') - \frac{\rho}{\rho'}\boldsymbol\nabla'\cdot(P'\uh) - \frac{\rho}{\rho'}\boldsymbol\nabla'\cdot(P_M\uh) + \rho(\uh\cdot\uh')(\dvp) \\ \label{ns}
	& -(\uh'\cdot\ua)\boldsymbol\nabla\cdot(\rho\ua) - \rho(\uh'\cdot\ua)(\da) -\rho(\uh\cdot\ua')(\dap) - \rho(\uh\cdot\ua')(\dap) + d_k + f_k,
\end{align}
\begin{align}\nonumber
	\partial_t(\rho\ua\cdot\ua') =& - \boldsymbol\nabla\cdot[(\ua\cdot\ua')\rho\uh] + \boldsymbol\nabla\cdot[(\uh\cdot\ua')\rho\ua] - \boldsymbol\nabla'\cdot[(\ua'\cdot\ua)\rho\uh'] + \boldsymbol\nabla'\cdot[(\uh'\cdot\ua)\rho\ua'] \\ \nonumber
	& -\frac{1}{2}\rho(\ua\cdot\ua')(\dv) - \frac{1}{2}\rho(\ua\cdot\ua')(\dvp) + \rho(\ua\cdot\ua')(\dvp) \\ \label{in}
	& -(\uh\cdot\ua')\boldsymbol\nabla\cdot(\rho\ua) - \rho(\uh'\cdot\ua)(\dap) + d_m,
\end{align}
\begin{align}\label{co}
	\partial_t(\rho e') =& - \boldsymbol\nabla'\cdot(\rho e'\uh') - \boldsymbol\nabla'\cdot(\rho e'\uh)  - \boldsymbol\nabla'\cdot(P\uh')  + \rho e'(\dvp),
\end{align}
where we have defined the dissipation and forcing correlators as,
\begin{align}
	d_k &= \langle\textbf{D}_k\cdot\uh'+\textbf{D}_k'\cdot\uh\rangle, \\
	f_k &= \langle\textbf{F}_k\cdot\uh'+\textbf{F}_k'\cdot\uh\rangle, \\
	d_m &= \langle\textbf{D}_m\cdot\ua'+\textbf{D}_m'\cdot\ua\rangle.
\end{align}
For a proper analysis of each term involved in the derivation, let us consider those terms that involve the divergence of third- and fourth-order variables (in particular those appearing in the first lines of Eqs.~\ref{ns}, \ref{in} and \ref{co}). After straightforward ordering and using the basic properties for homogeneous turbulence calculations \citep{Ba1953}, (i.e., $\langle\boldsymbol\nabla'\cdot(~)\rangle=\boldsymbol\nabla_r\cdot\langle\rangle$, $\langle\boldsymbol\nabla\cdot(~)\rangle=-\boldsymbol\nabla_r\cdot\langle\rangle$ and $\langle \alpha \rangle = \langle\alpha' \rangle$, with $\alpha$ any scalar function), the terms that involve the divergence of third- and fourth-order variables can be grouped as,
\begin{align}\label{re}
	\boldsymbol\nabla_r\cdot\langle R_E\uh - R_H\ua - R_E\uh' + R_H\ua' \rangle &= \boldsymbol\nabla_r\cdot\langle -R_E\delta\uh + R_H\delta\ua \rangle,
\end{align}
where we have introduced the usual increments definition $\delta\alpha\equiv\alpha'-\alpha$. Considering the prime version of Eq.~\eqref{re}, it can be shown that,
\begin{align}
	-\boldsymbol\nabla_r\cdot\langle (R_E+R_E')\delta\uh \rangle &= \frac{1}{2}\boldsymbol\nabla_r\cdot\langle [\delta(\rho\uh)\cdot\delta\uh+\delta(\rho\ua)\cdot\delta\ua + 2\delta e\delta\rho\big]\delta\uh\rangle - \langle E'(\dv)\rangle - \langle E(\dvp)\rangle, \\
	\boldsymbol\nabla_r\cdot\langle (R_H+R_H')\delta\ua \rangle &= -\frac{1}{2}\boldsymbol\nabla_r\cdot\langle [\delta(\rho\uh)\cdot\delta\ua+\delta(\rho\ua)\cdot\delta\uh\big]\delta\ua \rangle + \langle H'(\da) \rangle + \langle H(\dap) \rangle,
\end{align}
where we can identify two different types of terms, the so-called \textit{flux} and \textit{source} terms \citep{Ga2011}. The flux terms are those that can be written as divergence of the product of different increments, and they characterize the nonlinear cascade of energy. The source terms are those proportional to the local divergence of vectorial fields, such as $\uh$, $\ua$ or $\rho\uh$.  It is worth mentioning that these source terms are related to the dilatation (or contraction) of the plasma and are not present in IMHD turbulence~\citep{Ga2011,B2013}.

Defining the source terms $S_1$ as those proportional to $\dv$ and $S_2$ as those proportional to $\da$, Eqs. \eqref{ns}, \eqref{in} and \eqref{co} (and their prime version), can be grouped as,
\begin{align}
	S_1 &= \langle[R_E'-\frac{1}{2}(R_B'+R_B)](\dv)\rangle + \langle[R_E-\frac{1}{2}(R_B+R_B')](\dvp)\rangle, \\
	S_2 &= \langle[(R_H-R_H')-\bar{\rho}(\uh'\cdot\ua)](\da)\rangle+\langle[(R_H'-R_H)-\bar{\rho}(\uh\cdot\ua')](\dap)\rangle,
\end{align}
where $R_B\equiv\rho\ua\cdot\ua'/2$ is the magnetic energy density and $\bar{\alpha}=(\alpha'+\alpha)/2$ is the local mean value of a variable $\alpha$.

Finally, let us consider those terms that involve the magnetic and the plasma pressure. In particular, these terms can be cast as,
\begin{align}
	-\langle\boldsymbol\nabla\cdot(P\uh'+P_M\uh')\rangle -2\langle\boldsymbol\nabla'\cdot(P\uh')\rangle &= \langle(P_M-P)(\dvp)\rangle, \\
	-\langle\frac{\rho}{\rho'}\boldsymbol\nabla'\cdot(P'+P_M')\uh)\rangle &=  \langle \big(e'+ \frac{u_\text{A}^{'2}}{2}\big)\big[\boldsymbol\nabla\cdot(\rho\uh)\big] \rangle - \langle\frac{{\beta'}^{-1}}{2}\boldsymbol\nabla'\cdot(e'\rho\uh) \rangle,
\end{align}
 {where $\beta\equiv 2c_s^2/u_\text{A}^2$}. We can identify two types of source terms, those proportional to $\dv$ and those proportional to $\boldsymbol\nabla\cdot(\rho\uh)$. In addition, a new type of \textit{mixed} term appears, which is proportional to the divergence of a vector evaluated at the point $\textbf{x}$ multiplied by a variable taken at the point $\textbf{x}'$ and which, \textit{a priori}, cannot be transformed into a flux or a source term. In Section~\ref{discussion}, we analyze in detail each type of term and its role in the nonlinear dynamics of CMHD turbulence.

Combining the different terms, an expression for $\partial_t\langle R_E+R_E'\rangle$ can be obtained
\begin{align}\nonumber
	\partial_t \langle R_E + R_E'\rangle =&~ \frac{1}{2}\boldsymbol\nabla_r\cdot\big\langle [(\delta(\rho\uh)\cdot\delta\uh+\delta(\rho\ua)\cdot\delta\ua + 2\delta e\delta\rho\big]\delta\uh - [\delta(\rho\uh)\cdot\delta\ua+\delta\uh\cdot\delta(\rho\ua)]\delta\ua \big\rangle \\ \nonumber
	&+\langle[R_E'-\frac{1}{2}(R_B'+R_B)-E'+\frac{P_M'-P'}{2}](\dv)+[R_E-\frac{1}{2}(R_B+R_B')-E+\frac{P_M-P}{2}](\dvp)\rangle \\ \nonumber
	&+\langle[(R_H-R_H')-\bar{\rho}(\uh'\cdot\ua)+H'](\da)+[(R_H'-R_H)-\bar{\rho}(\uh\cdot\ua')+H](\dap)\rangle \\\nonumber
	& +\frac{1}{2}\langle\big(e'+\frac{u_\text{A}}{2}^{'2}\big)\big[\boldsymbol\nabla\cdot(\rho\uh)\big] + (e+\frac{u_\text{A}}{2}^2\big)\big[\boldsymbol\nabla'\cdot(\rho'\uh')\big]\rangle \\ \label{exact0}
	&-\frac{1}{2}\langle\beta^{-1'}\boldsymbol\nabla'\cdot(e'\rho\uh) + \beta^{-1}\boldsymbol\nabla\cdot(e\rho'\uh') \rangle + \mathcal{F} + \mathcal{D},
\end{align}
where $\mathcal{D}\equiv d_k+d_m$ and $\mathcal{F}\equiv f_m$.

Following the same assumptions used in fully developed turbulence (infinite kinetic and magnetic Reynolds numbers, a stationary state with a balance between forcing and dissipation \citep{Ga2011,B2013,A2016b,A2016c}), the exact law for CMHD turbulence can be obtained,
\begin{align}\nonumber
	-2\varepsilon =& \frac{1}{2}\boldsymbol\nabla_r\cdot\big\langle [(\delta(\rho\uh)\cdot\delta\uh+\delta(\rho\ua)\cdot\delta\ua + 2\delta e\delta\rho\big]\delta\uh - [\delta(\rho\uh)\cdot\delta\ua+\delta\uh\cdot\delta(\rho\ua)]\delta\ua \big\rangle \\ \nonumber
&+\langle[R_E'-\frac{1}{2}(R_B'+R_B)-E'+\frac{P_M'-P'}{2}](\dv)+[R_E-\frac{1}{2}(R_B+R_B')-E+\frac{P_M-P}{2}](\dvp)\rangle \\ \nonumber
&+\langle[(R_H-R_H')-\bar{\rho}(\uh'\cdot\ua)+H'](\da)+[(R_H'-R_H)-\bar{\rho}(\uh\cdot\ua')+H](\dap)\rangle \\\nonumber
& +\frac{1}{2}\langle\big(e'+\frac{u_\text{A}}{2}^{'2}\big)\big[\boldsymbol\nabla\cdot(\rho\uh)\big]+\big(e+\frac{u_\text{A}}{2}^2\big)\big[\boldsymbol\nabla'\cdot(\rho'\uh')\big]\rangle \\ \label{exactlaw}
&-\frac{1}{2}\langle\beta^{-1'}\boldsymbol\nabla'\cdot(e'\rho\uh) + \beta^{-1}\boldsymbol\nabla\cdot(e\rho'\uh') \rangle
\end{align}
where $\varepsilon=2\mathcal{F}$ is the mean dissipation or cascade rate. We recall that the exact law \eqref{exactlaw} is valid only in the inertial range and, therefore, it is independent of the dissipation mechanism present in the plasma (assuming that the dissipation acts only in the smallest scales of the system). {Introducing the compressible Elsasser variables, it can be shown that Eq.~\eqref{exactlaw} is equivalent to Eq. (19) in \citet{B2013}.}

\section{Discussion}\label{discussion}

In this Section, we analyze in detail each type of term involved in the exact law of isothermal CMHD turbulence derived in the previous Section and examine its role in the nonlinear energy transfer. This analysis should help when applying to exact law to {\it in situ} observations in space plasmas and to numerical simulation results.

\subsection{The flux terms}\label{flux_sec}

The flux terms are the usual terms present in exact laws of incompressible  turbulence, which are written as the global divergence of products of increments of the different variables. The compressible flux terms in Eq.~\eqref{exactlaw} can be grouped as,
\begin{align}\label{flux}
	\textbf{F}_\text{C} &= \textbf{F}_1+\textbf{F}_2 = \big\langle[(\delta(\rho\uh)\cdot\delta\uh+\delta(\rho\ua)\cdot\delta\ua\big]\delta\uh - [\delta(\rho\uh)\cdot\delta\ua+\delta\uh\cdot\delta(\rho\ua)]\delta\ua\big\rangle + 2\langle\delta e\delta\rho\delta\uh\rangle,
\end{align}
where $\textbf{F}_1$ can be identified as the Yaglom-like term \citep{MY1975} and $\textbf{F}_\text{2}$ corresponds to a new purely compressible flux term,
\begin{align}\label{fluxes}
	\textbf{F}_1 &= \big\langle[(\delta(\rho\uh)\cdot\delta\uh+\delta(\rho\ua)\cdot\delta\ua\big]\delta\uh - [\delta(\rho\uh)\cdot\delta\ua+\delta\uh\cdot\delta(\rho\ua)]\delta\ua\big\rangle,  \\
	\textbf{F}_2  &= 2\langle\delta e\delta\rho\delta\uh\rangle.
\end{align}
In the incompressibility limit (i.e., $\rho\rightarrow\rho_0=1$), the term \textbf{F}$_2$ tends to zero and the Yaglom-like flux term \textbf{F}$_1$ tends to the incompressible flux term \textbf{F}$_I=\big\langle [(\delta\uh)^2+(\delta\textbf{B})^2]\delta\uh - 2(\delta\uh\cdot\delta\textbf{B})\delta\textbf{B} \big\rangle$~\citep{P1998b}. Since all the source terms vanish in this limit, Eqs. \eqref{exactlaw} reduces to
\begin{align}\label{exact_mhd}
	-2\varepsilon_I &= \frac{1}{2}\boldsymbol\nabla_r\cdot\big\langle [(\delta\uh)^2+(\delta\textbf{B})^2]\delta\uh - 2(\delta\uh\cdot\delta\textbf{B})\delta\textbf{B} \big\rangle,
\end{align}
Equation \eqref{exact_mhd} corresponds to the exact law of IMHD turbulence \citep[see, e.g.][]{P1998a,P1998b}. Here \textbf{B} is expressed in {Alfv\'en} velocity units and $\varepsilon_I$ is the incompressible energy cascade rate. It is worth mentioning that \textbf{F}$_I$ depends only on the increments of the magnetic (and velocity) field, although the total magnetic field has been considered in the derivation. We will return to this point in Section~\ref{mean_sec}.

Another important remark that can be made here is that the total compressible flux term \textbf{F}$_\text{C}$ (Eq.~\ref{flux}) is a combination of fourth- and third-order terms (i.e., \textbf{F}$_1$ and \textbf{F}$_2$, respectively). This is a major difference with respect to the incompressible case, where \textbf{F}$_I$ (Eq.~\ref{exact_mhd}) involves only third-order correlations. The presence of fourth-order correlation function in the term \textbf{F}$_1$ is a direct consequence of the {\it total} energy in the CMHD model that combines a third-order kinetic energy and a second-order magnetic and internal compressible energies, and therefore, the correlation function $R_E$ must reflect this fact.

\subsection{The source and hybrid terms}\label{source_sec}

As we mentioned in Section \ref{derivation}, by definition the source terms are those proportional to the divergence of the vectorial fields $\uh$, $\ua$ or $\rho\uh$. In principle, the source-like terms in Eq.~\eqref{exactlaw} are,
\begin{align}\nonumber
	\text{S} =&~ \langle[R_E'-\frac{1}{2}(R_B'+R_B)-E'+\frac{P_M'-P'}{2}](\dv)+[R_E-\frac{1}{2}(R_B+R_B')-E+\frac{P_M-P}{2}](\dvp)\rangle \\\nonumber
	&+\langle[(R_H-R_H')-\bar{\rho}(\uh'\cdot\ua)+H'](\da)+[(R_H'-R_H)-\bar{\rho}(\uh\cdot\ua')+H](\dap)\rangle \\ \label{source}
	& +\frac{1}{2}\langle\big(e'+\frac{u_\text{A}}{2}^{'2}\big)\big[\boldsymbol\nabla\cdot(\rho\uh)\big]\rangle + \frac{1}{2}\langle\big(e+\frac{u_\text{A}}{2}^2\big)\big[\boldsymbol\nabla'\cdot(\rho'\uh')\big]\rangle.
\end{align}
Note that the structure of each of these source terms is the same: a scalar function $f$ multiplied by the divergence of a given vector \textbf{G}, plus its prime version. However, depending on the spatial dependence of the scalar function $f$, two situations can be faced. First, if the scalar function depends only on single point in space \textbf{x}, i.e. $f=f(\textbf{x})$, it is straightforward to demonstrate that,
\begin{align}\label{prop}
	\langle f'\boldsymbol\nabla\cdot\textbf{G} + f\boldsymbol\nabla'\cdot\textbf{G}'  \rangle &= \boldsymbol\nabla_r\cdot\langle 2\bar{f}\delta\textbf{G}\rangle.
\end{align}
In this case, the source terms of Eq.~\eqref{source} can be converted into flux-like terms (and vice versa). Here we refer to these terms as \textit{hybrid} terms. The terms in Eq.~\eqref{source} that satisfy this condition are
\begin{align}\nonumber
	\text{S}_\text{H} =&~ \langle\big(\frac{P_M'-P'}{2}-E'\big)(\dv)+\big(\frac{P_M-P}{2}-E\big)(\dvp)\rangle  + \langle H'(\da)+H(\dap)\rangle\\ \label{hybrid_s}
	& +\frac{1}{2}\langle\big(e'+\frac{u_\text{A}}{2}^{'2}\big)\big[\boldsymbol\nabla\cdot(\rho\uh)\big]+\big(e+\frac{u_\text{A}}{2}^2\big)\big[\boldsymbol\nabla'\cdot(\rho'\uh')\big]\rangle,
\end{align}
These terms can be converted into flux-like terms using Eq.~\eqref{prop}
\begin{align}\label{hybrid_f}
	\textbf{F}_\text{H} =&~ \boldsymbol\nabla_r\cdot\langle\big(\bar{P}_M-\bar{P}-2\bar{E}\big)\delta\uh + 2\bar{H}\delta\ua + \big(\bar{e}+\frac{\bar{u}_\text{A}}{2}^{2}\big)\delta(\rho\uh)\rangle.
\end{align}
We emphasize that these flux-like terms \eqref{hybrid_f} are not mere products of increments, as the typical flux terms \eqref{flux}, but rather involve averaged quantities between the points $\textbf{x}$ and $\textbf{x}'$.

Second, if the scalar function $f$ depends on two points $\textbf{x}$ and $\textbf{x}'$, i.e. $f=f(\textbf{x,x'})$, then Eq.~\eqref{prop} is not valid. This applies to all the terms in Eq.~\eqref{source} that involve two-points correlation functions, namely $R_E$, $R_H$ and $R_B$ (and their prime version). These are the true compressible source terms $S_\text{C}$.
\begin{align}\nonumber
	\text{S}_\text{C} =&\langle[R_E'-\frac{1}{2}(R_B'+R_B)](\dv)+[R_E-\frac{1}{2}(R_B+R_B')](\dvp)\rangle \\\label{source_c}
	&+\langle[(R_H-R_H')-\bar{\rho}(\uh'\cdot\ua)](\da)+[(R_H'-R_H)-\bar{\rho}(\uh\cdot\ua')](\dap)\rangle.
\end{align}

\subsection{The $\beta$-dependent terms}\label{mixed_sec}

The $\beta$-dependent terms in Eq.~\eqref{exactlaw} cannot not {\it a priori} written as flux or source terms,
\begin{align}\label{mixed}
	\text{M}_\beta &= -\frac{1}{2}\langle\beta^{-1'}\boldsymbol\nabla'\cdot(e'\rho\uh) + \beta^{-1}\boldsymbol\nabla\cdot(e\rho'\uh') \rangle.
\end{align}
Note that this new type of term is present only CMHD and has no equivalence in compressible HD equations. In fact, these terms stem from the magnetic pressure gradient term in the momentum Eq.~\eqref{1}. Assuming statistical stationarity of the $\beta$ parameter, the terms in Eq.~\eqref{mixed} can be converted into a hybrid (flux or source) term \citep[see also][]{B2016a},
\begin{align}\nonumber\label{eq_phi}
	\text{M}_\beta &= -\frac{1}{2}\langle\boldsymbol\nabla'\cdot(\beta^{-1'}e'\rho\uh) + \boldsymbol\nabla\cdot(\beta^{-1}e\rho'\uh') \rangle \\
	&= \boldsymbol\nabla_r\cdot\langle\overline{\beta^{-1}e}\delta(\rho\uh)\big\rangle = \frac{1}{2}\langle\beta^{'-1}e'\big[\boldsymbol\nabla\cdot(\rho\uh)\big]+\beta^{-1}e\big[\boldsymbol\nabla'\cdot(\rho'\uh')\big]\rangle.
\end{align}

To summarize our results, equation \eqref{exactlaw} can be schematically written as,
\begin{equation}
	-2\varepsilon=\frac{1}{2}\boldsymbol\nabla_r\cdot\textbf{F}_\text{C}+\text{S}_\text{C}+\text{S}_\text{H}+\text{M}_\beta,
\end{equation}
where $\textbf{F}_\text{C}$, $\text{S}_\text{C}$, $\text{S}_\text{H}$ and M$_\beta$ represent respectively the pure compressible flux, source, hybrid and $\beta$-dependent terms. This schematic representation reflects the true nature of each term in the exact law of CMHD turbulence. Note that in the observational works of \citet{B2016a} and \citet{H2017,H2017b}, the terms $\text{F}_\text{C}$, $\text{M}_\beta$ (under the assumption of stationary $\beta$) and part of the terms $\text{S}_\text{H}$ were considered in the evaluation of the energy cascade rate, the remaining terms were considered as source terms and assumed to be subdominant.

\subsection{Comparison between CMHD and IMHD: role of the background magnetic field ${\bf B}_0$}\label{mean_sec}

In Section \ref{derivation}, we have derived an exact law for fully developed CMHD isothermal turbulence. Our theoretical results in Eq.~\eqref{exact0} involve a second- and third-order dynamical equation for the energy correlation function $R_E({\bf x},{\bf x}')$ expressed as a function of second-, third- and fourth-order correlation functions of the turbulent fields. In contrast, the vKH equation of IMHD turbulence is a second-order equation for the correlation energy function expressed as a function of third- and second-order correlation terms \citep[e.g.][]{P1998a}. Another significant difference between CMHD and IMHD is that the exact law in IMHD (Eq.~\ref{exact_mhd}) has no dependence on the (constant) background field ${\bf B}_0$, and rather depends only on the increments on ${\bf B}$ and ${\bf v}$. This fact is somehow puzzling since IMHD turbulence is anisotropic in nature, and the background field ${\bf B}_0$ is an essential ingredient in its dynamics \citep[see, e.g.][]{Mo1981,Sh1983,H1984,G1995,Mo1995,M1996,G1999,BN2001,S2016}. However, the dependence on ${\bf B}_0$ in IMHD can be recovered when the derivation is pushed to the next order. {This has been done by \citet{W2012} who showed that the third order dynamical equation depends explicitly on ${\bf B}_0$ \citep[see also, ][]{O2013}.}

This fundamental difference in order between the vKH equations in IMHD and CMHD should be kept in mind when comparing the results about the cascade rate estimated from the two models in numerical simulations or spacecraft observations. For instance, the amplification of the cascade rate reported in spacecraft observations in the solar wind and magnetosheath using the exact law of CMHD \citep{B2016a,H2017,H2017b} is certainly due to density fluctuations {\it but also to their coupling} with ${\bf B}_0$ (a similar role of $\rho_0$ has been pointed out in \citet{H2017}). The corresponding nonlinear terms have no counterpart in the IMHD exact law. This suggests that the energy cascade rates given in previous studies that used the IMHD model in the solar may have been underestimated. The original works of \citet{B2016a,H2017,H2017b} based on the exact law of CMHD have improved those estimations, but since some of the terms discussed here (some part of the hybrid terms in Eq.~\ref{hybrid_f} and the pure source ones in Eq.~\ref{source_c}) were not included in those works, more complete studies should be performed to assess the contribution of each missing term to the total energy cascade rate. Those studies should use realistic 3D CMHD simulations and possibly multi-spacecraft data that would allow one to evaluate the importance of the source terms.

A final remark concerns the total energy of the CMHD system (which is a conserved quantity) that has been considered to derive the corresponding exact law \eqref{exactlaw}. As mentioned above, the fourth-order terms appearing in Eq.~\eqref{exact0} (which has no counterpart in IMHD) stems from the kinetic energy $\rho v^2/2$ that is a third order, while it is second-order in IMHD. Therefore, one may ask whether a {reduced} form of Eq.~\eqref{exact0} derived for the fluctuating (second-order) energy, i.e., $E_1(\textbf{x}) = \rho_0\uh_1\cdot\uh_1/2+{\bf b}_1\cdot{\bf b}_1 + \rho_1 e_1$ (although it is not an invariant of motion), would have fourth-order dependence and whether the zeroth order variable ${\bf B}_0$ and $\rho_0$ will appear explicitly in the result. The calculation (not shown here) show no fourth-order term as in Eq.~\eqref{exact0}, a situation similar to IMHD, but the background magnetic field ${\bf B}_0$ present in the result, which again contrasts with the IMHD model. Upcoming numerical simulations results will shed light on the role of the background quantities and their coupling to turbulent fluctuations in CMHD turbulence.

\section{Conclusions}\label{conclusions}

We  have derived the exact law for fully developed homogeneous CMHD turbulence, without the assumption of isotropy. For an isothermal plasma, we have presented the exact law as a function of the plasma velocity field, the compressible Alfv\'en velocity and the plasma density. In particular, our theoretical results show four different categories of terms that are involved in the nonlinear energy cascade  in the inertial range, which are the hybrid and $\beta$-dependent terms, in addition to well-known flux and source terms. The hybrid terms offer the freedom to be written either as flux terms  or source terms. However, as a flux-like terms (see Eq.~\ref{hybrid_f}) they cannot be written as the product of increments, as the usual flux terms (see Eq.~\ref{flux}). A $\beta$-dependent term (already reported as a flux term in \citep{B2013}), which has no counterpart in compressible HD turbulence, cannot in general be expressed  either as flux or source term. We compared the structure of the vKH equations in CMHD and IMHD and showed that the former has a higher (fourth) order dependence on the turbulent fields than the latter. Furthermore, and unlike the exact law in IMHD, the vKH equation in CMHD was shown to depend explicitly on the background magnetic field ${\bf B}_0$  that is coupled to the fluctuating fields. This is clearly a major difference with the exact law in IMHD, which may explain the amplification of the cascade rate reported in the solar wind and magnetosheath given by the CMHD model with respect to the estimates from the IMHD model~\citep{B2016a,H2017,H2017b}. Three dimensional numerical simulations of CMHD are needed to better quantify the contribution of each term to the total cascade rate and its dependence on the background magnetic field ${\bf B}_0$.

\section*{Acknowledgments}

NA and FS thank S. Banerjee, S. Galtier and L. Z. Hadid for useful discussions. NA is supported through an \'Ecole Polytechnique Postdoctoral Fellowship and by LABEX Plas@Par through a grant managed by the Agence Nationale de la Recherche (ANR), as part of the program “Investissements d’Avenir” under the reference ANR-11-IDEX-0004–02.  FS and NA acknowledge financial support from Programme National Soleil-Terre (PNST).

\bibliographystyle{apsrev4-1}

%

\end{document}